\documentclass[12pt]{article}\usepackage[hyperfootnotes=false]                  {hyperref}
\usepackage{epsfig}
\usepackage{float}

\usepackage{caption}

\usepackage{amsmath}
\usepackage{amssymb}
\usepackage{graphicx}
\newlength{\abstractwidth}
\renewcommand{\thefootnote}{\fnsymbol{footnote}}
\renewcommand{\thanks}[1]{\footnote{#1}} 
\newcommand{\starttext}{
\setcounter{footnote}{0}
\renewcommand{\thefootnote}{\arabic{footnote}}}

\newcommand{\be}{\begin{equation}}
\newcommand{\bea}{\begin{eqnarray}}
\newcommand{\eea}{\end{eqnarray}}
\newcommand{\beq}{\begin{equation}}
\newcommand{\ee}{\end{equation}}

\def\eq{&=&}

\def\simleq{\; \raise0.3ex\hbox{$<$\kern-0.75em
\raise-1.1ex\hbox{$\sim$}}\; }
\def\simgeq{\; \raise0.3ex\hbox{$>$\kern-0.75em
\raise-1.1ex\hbox{$\sim$}}\; }

\def\bi{\begin{itemize}}
\def\ei{\end{itemize}}

\def\S{Schwarzschild}
\def\sc{\setcounter{equation}{0}}

\def\CC{{\cal{C}}}

\def\CL{{\cal{L}}}

\def\t{\tau}

\def\bn{\bigskip \noindent}

  \def\tb{\tilde{\beta}}
  \def\tt{\tilde{T}}

\makeatletter
\g@addto@macro\normalsize{%
  \setlength\abovedisplayskip{10pt}
  \setlength\belowdisplayskip{20pt}
  \setlength\abovedisplayshortskip{10pt}
  \setlength\belowdisplayshortskip{20pt}
}
\makeatother

\usepackage{color}

\begin{document}
  
\begin{titlepage}

\rightline{}
\bigskip
\bigskip\bigskip\bigskip\bigskip
\bigskip

\centerline{\Large \bf {  Falling Toward Charged Black Holes}}
\bn

\bigskip
\begin{center}
\bf   Adam R.~Brown$^1$, Hrant Gharibyan$^1$, Alexandre Streicher$^{1,2}$,  \\ Leonard Susskind$^1$, L\'arus Thorlacius$^{1,3}$, Ying Zhao$^1$ \rm

\bigskip
$^1$ Stanford Institute for Theoretical Physics and Department of Physics, \\
Stanford University,
Stanford, CA 94305-4060, USA \\

\bigskip
$^2$ Department of Physics, University of California, Santa Barbara, CA 93106, USA

\bigskip
$^3$ University of Iceland, Science Institute, Dunhaga 3, 107 Reykjavik, Iceland

and

The Oskar Klein Centre for Cosmoparticle Physics, Department of Physics, 
Stockholm University, AlbaNova, 106 91 Stockholm, Sweden

\end{center}

\bn

\begin{abstract}

The growth of the ``size" of operators is an important diagnostic of quantum chaos. 
In \cite{Susskind:2018tei} it was conjectured that the holographic dual of the size  is proportional to the average radial component of the momentum of the particle created by the operator. Thus the growth of operators  in the background of a black hole corresponds to the acceleration of the particle as it falls toward the horizon.

In this note we will use the momentum-size correspondence as a tool to study scrambling in the field of  a near-extremal charged black hole. The agreement with previous work  provides a non-trivial test of the momentum-size relation, as well as an explanation of a  
paradoxical  feature of scrambling previously  discovered by Leichenauer \cite{Leichenauer:2014nxa}. Naively Leichenauer's result says that only the non-extremal entropy participates in scrambling. The same feature is also present in the SYK model. 
 
 In this paper we find a quite different interpretation of Leichenauer's result which does not have to do with any decoupling of the extremal degrees of freedom. Instead it has to do with the buildup of momentum as a particle accelerates through the long throat of the Reissner-Nordstr{\"o}m geometry.  \\
 
 Version 3: in this version of the paper, the conjectured size-momentum relation has been significantly modified. Rather than the proportionality factor being constant, we now conjecture that it varies through the throat.  This result also agrees with forthcoming direct calculations in SYK.

 \bn

\medskip
\noindent
\end{abstract}

\end{titlepage}

\starttext \baselineskip=17.63pt \setcounter{footnote}{0}
\tableofcontents

\section{Two Puzzles }\label{intro}

All horizons are locally the same; namely they are  Rindler-like%
\footnote{Extremal black holes are an exception. In this paper we consider the limit in which the non-extremality parameter $(r_+ - r_-)/r_+ $ is arbitrary but fixed as $r_+$ becomes large. In this limit the horizon is Rindler-like.}.
 Therefore one might expect their properties as scramblers and 
 complexifiers to be universal. For example,  the rate of growth 
 of complexity  for all neutral static black holes scales as\footnote{In this paper there are many undetermined numerical 
 coefficients, partly because they depend on precise definitions 
 that we leave unspecified. For example the definition of  complexity is ambiguous up to a numerical factor.  Our 
 convention will be to use the symbol $\sim$ to mean ``equal 
 up to a numerical factor''. By and large additive numerical ambiguities of order unity, such as in the definition of scrambling time, will be ignored.}

\be 
\frac{d\CC}{dt} \sim \frac{S}{R_s} ,
\label{C-rate}
\ee
where $S$ and $R_s $  are the entropy and Schwarzschild radius of the black hole \cite{Brown:2015bva}. Similarly there is a universal formula for the scrambling time \cite{Maldacena:2015waa},
\be 
t_* = \frac{\beta}{2\pi} \log{\frac{S}{\delta S}},
\label{scrambling}
\ee
where $\delta S = \delta E/T$, and $\delta E$ is the energy of the initial perturbation.

It is therefore surprising that charged black holes  behave differently. For charged black holes,  Eqs.~\ref{C-rate} and \ref{scrambling} are modified to \cite{Brown:2015bva,Leichenauer:2014nxa}
\begin{eqnarray}
\frac{d\CC}{dt} & \sim &\frac{S-S_0}{R_s} \label{C-rate-charged} \\
t_* &=& \frac{\beta}{2\pi} \log{\left( \frac{S-S_0}{\delta S} \right)  } . \label{scrambling-charged}
\end{eqnarray}
Here $R_s$ is the area-radius of the horizon (otherwise known as $r_+$) and $S_0$ is the entropy of the extremal black hole with the same charge. 

\bn

If instead of using \S \ time $t$ we use dimensionless Rindler time, defined by
\be 
\tau = \frac{2\pi t}{\beta},
\ee
then Eqs.~\ref{scrambling} and \ref{scrambling-charged} take an especially simple form: for uncharged black holes
\be 
\tau_* = \log\frac{S}{\delta S} \ ,
\label{tau-scrmbl-1}
\ee
and for charged black holes, described by the Reissner-Nordstr{\"o}m (RN) metric,
\be 
\tau_* = \log\frac{(S-S_0)}{\delta S} .
\label{tau-scrmbl-2}
\ee
\bn

Exactly the same features, namely Eqs.~\ref{C-rate-charged} and \ref{scrambling-charged}, also hold for the black holes described by the SYK model. Because the SYK model is a precise quantum mechanical system, we can hope to track down the microscopic origin of this behavior.

 A simple explanation would be that the extremal degrees of freedom are somehow decoupled from the chaotic behavior,  leaving  only the non-extremal component  to actively ``compute".   But given the fact that all horizons are Rindler-like, it is hard to understand why this should be so. 

We will propose an interpretation of Eqs.~\ref{C-rate-charged} and \ref{scrambling-charged} that has nothing to do with any decoupling of extremal degrees of freedom. Horizons  do indeed have universal computational properties in which all $S$ degrees of freedom  actively compute. Neutral and charged black hole horizons compute  in exactly the same way.  

For simplicity, in this paper we will work with asymptotically flat black holes. Our results would apply equally to black holes of small or intermediate size in AdS. 

\subsection{Complexity Growth}

The explanation of \ref{C-rate-charged}  for the rate of complexity growth is simple. For near-extremal black holes, the entropy above extremality is linear in the temperature $T$
\be 
\frac{S-S_0}{S} \sim r_+T,
\label{DSoverS}
\ee
where $r_+$ is the area-radius of the outer horizon.  Thus we may write \ref{C-rate-charged} in the form,
\be 
\frac{d\CC}{dt} \sim TS \, .
\label{C-rate-all}
\ee
We can get more insight into the meaning of \ref{C-rate-all} by replacing the usual Schwarzschild time $t$ by the dimensionless Rindler time $\t = \frac{2\pi t}{\beta}$   (i.e.~the hyperbolic angle), which gives
\be 
\boxed{\frac{d\CC}{d\t} \sim S} 
\label{universal}
\ee
Equation \ref{universal} expresses a universal property of horizons, charged and uncharged: they all compute at a rate $\sim$  one gate per Rindler time per bit of entropy. All degrees of freedom participate; extremal degrees of freedom do not decouple.

\subsection{Scrambling}
The explanation of \ref{scrambling-charged} is not so simple. It will occupy the rest of the paper. Here is a quick summary of the method we will employ:

The basic tool is the size-momentum relation introduced in \cite{Susskind:2018tei}. 
A perturbation can be applied to a black hole by acting with an operator $W$. With time the operator evolves to the ``precursor" 
$W(t)= e^{iHt}We^{-iHt}$. In a holographic theory an operator like $W(t)$ can be characterized by its size---a measure of the average number of fundamental degrees of freedom making up the operator 
(see \cite{Roberts:2018mnp} and references therein)\footnote{The definition and calculations of size in \cite{Roberts:2018mnp} are appropriate to the infinite temperature limit. In this paper our definition of `size' involves a thermal averaging. See Sec.~\ref{Sec: SYK}}. Initially this number may be small, of order unity\footnote{It is expected that a simple operator such as an SYK fermion has a size equal to $1$.}. Subsequently the size grows, reaching its saturation value at the scrambling time. The saturation value is the entropy of the black hole.

If we denote the size at time $t$ by $s(t)$ then the scrambling time is determined from,
\be 
s(t_*) \sim S.
\label{s=S}
\ee

The action of the operator $W$ is to create a particle at the boundary  (to be defined) of the black hole geometry. The particle wave packet can be tracked as it falls toward the horizon.
The size-momentum relation is a duality between the  size of the precursor and the average radial momentum of the in-falling wave packet of the particle. By calculating the momentum we also calculate the size and this allows us to implement \ref{s=S} in an efficient manner.

In  \cite{Susskind:2018tei} this method was used to calculate the scrambling time for the case of a \S \ black hole in asymptotically flat space. Here we use the same method, with one new twist, to calculate the scrambling time for charged black holes in asymptotically flat space. The result not only agrees with \ref{scrambling-charged} but it does so while assuming that all degrees of freedom---not just the non-extremal degrees of freedom---participate in the scrambling dynamics.

\section{Geometry of  Charged Black Holes}
Consider the 3+1-dimensional Reissner-Nordstr{\"o}m  black hole, 
 \bea 
 ds^2 &=& -f(r)dt^2 + \frac{dr^2}{f(r)} + r^2 d\Omega^2 \cr \cr
 f(r) &=& \left(1-\frac{r_+}{r}\right) \left(1-\frac{r_-}{r}\right) .
 \label{RN-metric}
 \eea
 The inner (-) and outer (+) horizons are at  $r_{\pm} \equiv G M \pm \sqrt{G^2 M^2 - GQ^2}$, and 
the Hawking temperature is 
 \begin{equation}
 T = \frac{1}{4\pi}\left( \frac{r_+-r_-}{r_+^2} \right).
\end{equation}
All Reissner-Nordstr{\"o}m black holes must have $Q^2 \leq G M^2$ and $r_- \leq r_+$, and when these inequalities are saturated the black hole is said to be `extremal'. In this paper, we will be interested in near-extremal black holes, so that 
$r_+ - r_- \ll r_+$. In this limit, the temperature is small ($\beta \gg r_+$) and the near-horizon region develops a long `throat'. 

\subsection{The Geometry of the Throat} \label{sec:throatgeometry}
 
\begin{figure}[H]
\begin{center}
\includegraphics[scale=.5]{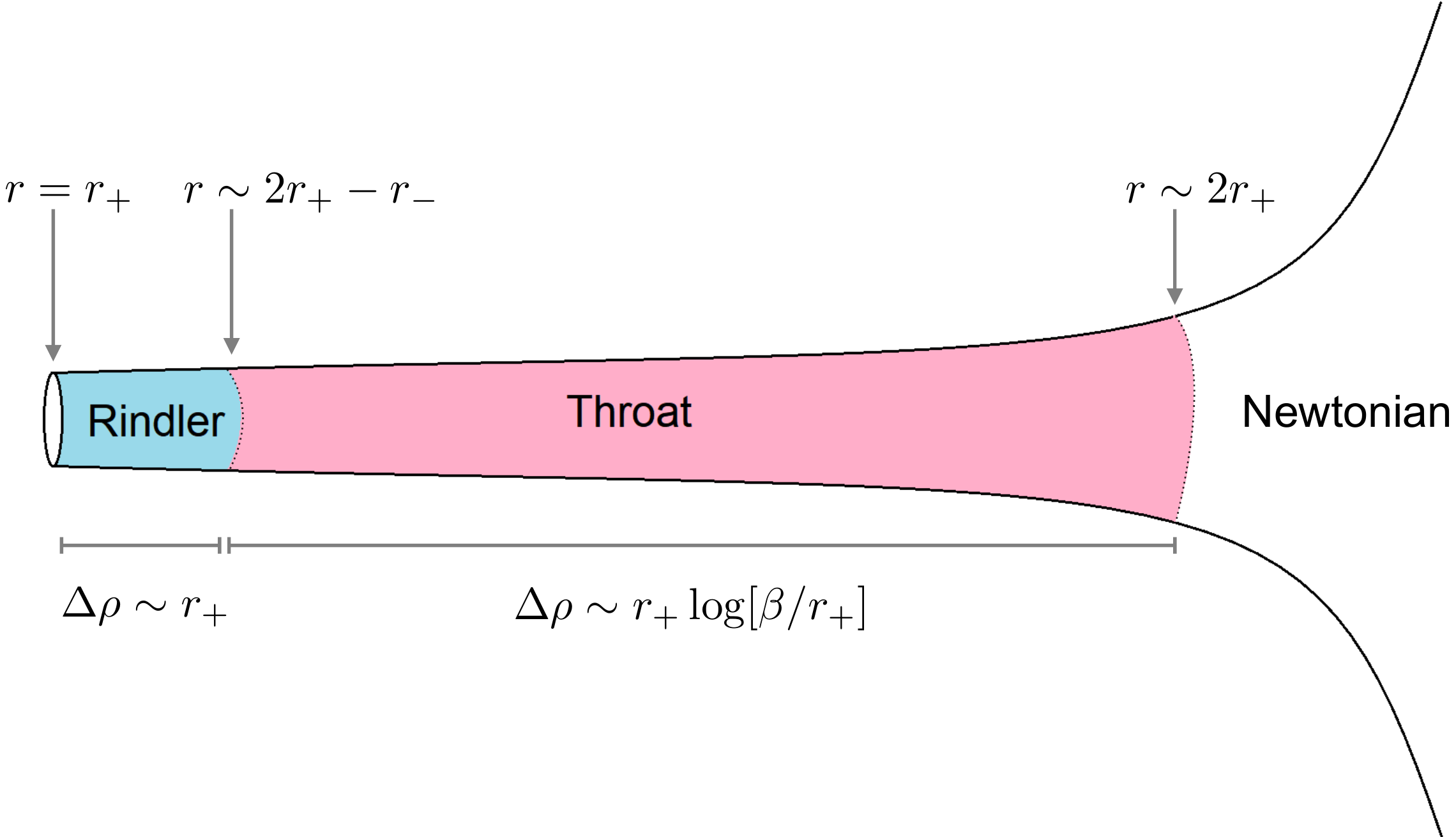}
\caption{The three regions outside a near-extremal charged black hole. Unlike for uncharged black holes, there is now a `throat' separating the Rindler and Newtonian regions.}
\label{throatla}
\end{center}
\end{figure}

 The exterior of a near-extremal black hole can be divided into three regions, as in Fig.~\ref{throatla}.

\begin{itemize}
\item The innermost region is the Rindler or near-horizon region where the geometry closely resembles the \S \ black hole with the same entropy. It is  defined by
\begin{eqnarray}
r_+ < &r&  \simleq \   \ 2r_+ - r_- \label{are} \\
0 < &\Delta \rho&  \simleq \ \  r_+ ,
\end{eqnarray}
where $\Delta \rho = \int dr/\sqrt{f(r)}$ is the proper distance from the outer horizon. The gravitational field (i.e.~the proper acceleration $\alpha = \partial_r \sqrt{f(r)}$ required to remain static at fixed $r$) grows rapidly near the horizon. While the quantity $(1 - \frac{r_+}{r})^{-1}$ varies significantly in the Rindler region, $(1 - \frac{r_-}{r})^{-1}$ is essentially constant. 

\item The next region out is the throat, defined by
\begin{eqnarray}
2r_+ - r_- \ \simleq &r&  \simleq \  \ 2r_+ \\
r_+  \ \simleq &\Delta \rho&  \simleq \ \   r_+  \log \left[ \frac{r_+}{r_+ - r_-} \right]  \ .
\end{eqnarray}
The throat is long and of approximately constant width (it resembles AdS$_2 \times S^2$) and the gravitational field is approximately constant. We have $(1 - \frac{r_-}{r})^{-1} \sim (1 - \frac{r_+}{r})^{-1}$ and both vary significantly through the throat. The throat ends at $r=2r_+$ which we will soon see is the location of a potential barrier which separates the throat from the Newtonian region.
The throat is unique to charged black holes; it is absent from the \S \ black hole. 

For most purposes the geometry in the throat can be approximated by the extremal geometry with $r_+ = r_-.$

(Later we will comment on the relation between near-extremal RN black holes and the SYK model. For now we note that the dynamical boundary of the AdS$_2$ dual of SYK should be identified with the end of the RN throat adjacent to the Newtonian region at $r=2r_+$.)

\item Outermost is the Newtonian region, where $(1 - \frac{r_-}{r})^{-1} \sim (1 - \frac{r_+}{r})^{-1} \sim 1$.
\end{itemize}

\subsection{The Black Hole Boundary}
Black holes in flat space  evaporate but the rate is extremely slow, and for many purposes can be ignored. The reason that the rate is so slow is that the black hole is effectively in a reflecting box. The box is the potential barrier that quanta experience as they try to escape the near-horizon region. Even the S-waves are inhibited by a barrier. For a \S \ black hole the barrier height is about equal to the temperature and there is significant leakage, but for a near-extremal RN black hole the barrier height is much higher than the temperature. The barrier provides a natural boundary of the black hole region and may be thought of as the holographic boundary in a quantum description.

For the spherically symmetric field mode, the potential barrier takes the form \cite{Susskind:2005js}
 \be
 V(r) = \frac{\partial_r(f^2)}{4r}.
 \ee 
For a RN black hole, Eq.~\ref{RN-metric} then gives
\be
V(r) =  \frac{\partial_r(f^2)}{4r}  = \frac{r_+(r-r_+)^3}{r^6}.
\ee
 The barrier is well localized at the end of the throat near where it meets the Newtonian region. The width in proper distance is of order $r_+$ and for near-extremal RN it is much narrower than the length of the throat.
 
 The potential at the top  of the barrier is,
 \be 
 V_\textrm{top} = \left( \frac{1}{8 r_+}  \right)^2 \ . 
 \ee
(Note that the potential has units of momentum-squared. In rolling down the potential a massless particle would gain a momentum $\sqrt{V_\textrm{top}}$.) 

We can consider the top of the potential barrier to be the boundary of the black hole. It occurs at 
\be 
r = 2 r_+ \equiv r_b \ . 
\ee 
Here the subscript $b$ stands either for barrier or boundary.

It will be convenient to define a radial proper-length coordinate $\rho$ measured from the black hole boundary,
\be 
\rho = \int_r^{r_b}\frac{dr'}{\sqrt{f(r'})} . 
\ee

At   the boundary $\rho=0$ and at the beginning of the Rindler region $\rho = r_+ \log{(\beta/r_+)}$.

\bn

\section{Size-Momentum Relation}

By the `size' of a particle we mean the average number of elementary operators making up the precursor. In  \cite{Roberts:2018mnp} the average was over an infinite temperature ensemble, but in this paper we are interested in low temperatures. The study of low temperature size and operator growth is not well developed but we will assume that these concepts can be generalized appropriately. 

\bn

By the radial momentum of a particle we mean the momentum that an observer at rest would see as the particle passed her. We imagine a particle passing an observer suspended at some distance $\rho$ from the boundary. The momentum measured by that observer is called $P$. We will generally assume that it is relativistic. Classically it is easy to track since the mechanism for its growth is ordinary blue-shift. Quantum mechanically we will assume that the average momentum of the wave packet follows the classical trajectory.

In \cite{Susskind:2018tei} it was conjectured that the average momentum $P$ at time $t$ is dual to the size of the
precursor $W(t)$. By time we mean the ordinary Schwarzschild time, not the proper
time of the particle.

The conjecture that size and average momentum are dual to one another requires us to specify a proportionality factor  with dimensions of length.  In  \cite{Susskind:2018tei} (the case of \S \ black holes) the proportionality factor was taken to be the \S \ radius, or what is the same thing, the thermal length $\beta,$
\bea 
\textrm{size} &\sim& R_s P \cr 
&\sim& \beta P \ . 
\label{old-proposal}
\eea

Although this identification is adequate for estimating the scrambling properties of \S \  black holes, there are reasons to believe that the correct connection between size and momentum is more complicated, and that it involves not only the momentum but also the position of the particle.
To see why this is so let us first consider two situations involving  infalling particles at the same location, but of different momentum. If it is not obvious it will soon become clear that the operator that creates the higher momentum particle has the larger size.

But now consider the opposite case: two particles of the same momentum but at different radial locations. It is well known that the deeper into the bulk the particle is, the more complex the operator must be that creates it. Complexity and size are closely related and it should not be surprising that the operator creating the deeper particle has larger size. Therefore we should expect that size is a function of both the radial momentum and the radial location of  the particle created by $W(t).$ This will involve  the concept of a local energy scale that varies significantly  throughout the long throat  of the near-extremal RN black hole. For the \S \ black hole there is no throat.

\subsection{The Local Energy Scale} \label{sec:localenergyscale}

This paper is about near-extremal Reissner-Nordstr{\"o}m charged black holes in asymptotically flat space. It is well known that the throat-geometry of such black holes is $AdS(2) \times S(2).$
As with all AdS geometries, $AdS(2) $ has a scaling symmetry, in this case associated
with the long throat. To see this, we consider black holes with a fixed total charge $Q$,
{\it i.e.} we fix the ground state entropy 
$S_0= \frac{\pi r_e^2}{G} =  \pi Q^2 $.
At finite temperature the excess entropy above the extremal value is given by
\begin{align}
	S-S_0 =\ & \frac{\pi r_+^2}{G} -\pi Q^2\nonumber \\
	=\ & 4\pi S \frac{r_+}{\beta}\label{dof_T} \\
	\approx\ & 4\pi S_0 \frac{r_+}{\beta}\,,\nonumber 
\end{align}
and it follows that $d\log(S-S_0) \approx -d\log\frac{\beta}{r_+}$, where in the differential here 
we hold the ground state entropy fixed.
Earlier we saw that for a near-extremal black hole with $\beta\gg r_+$ the proper length of the 
throat is $\rho(T) \approx r_+\log\frac{\beta}{r_+}$, and we can thus reexpress \eqref{dof_T} as
\be
	d\log(S-S_0) = -d\left(\frac{\rho}{r_+}\right)\,. \label{dof_rho} 
\ee
As we lower the temperature, the throat gets longer. Assuming $T_1<T_2\ll\frac{1}{r_+}$, we have 
\be
\label{scale1}
	\frac{S(T_1)-S_0}{S(T_2)-S_0} = \exp\left(-\frac{\rho(T_1)-\rho(T_2)}{r_+}\right).
\ee
We see that each time the length of the throat increases by $r_+$, $S-S_0$ is decreased by a factor of
$e^{-1}$. 

We can understand this from another point of view. Consider a near-extremal black hole at some low 
temperature $T\ll\frac{1}{r_+}$. From the holographic bound \cite{Susskind:1998dq}, the number of degrees 
of freedom necessary to completely describe the region within radius $r$ is given by its transverse area 
$A(r) = 4\pi r^2$. We start from the outer end of the throat $r = 2r_+$, where the area equals $A_\textrm{max}$, 
and then the area decreases as we move into the throat. The pattern of decrease becomes clear when we 
use the variable $\Delta A(r)\equiv A(r)-A_0$ rather than the area itself. Here $A_0 \equiv 4 \pi G Q^2$ is the horizon area of 
the extremal black hole carrying the same charge. The decrease in the transverse area as we go into the 
throat is then given by
\be 
\label{scale2}
	\frac{\Delta A(\rho)}{\Delta A(0)} =  \exp{\left(-\frac{\rho}{r_+} \right)},
\ee
where $\rho(r)$ is the proper distance from the boundary.
The exponential behavior continues until we reach the Rindler region.
The area itself does not decrease very much along the throat since $4 \pi r_+^2 < A(r) < A_\textrm{max} = 16 \pi r_+^2$; but the difference in area varies a great deal since at low temperature $\Delta A(r_+) \ll \Delta A_\textrm{max}$. As we move in the radial direction inside the throat, the quantity 
that scales with a well-defined scaling dimension is $\Delta A$. This is to be compared with the vacuum AdS case, 
where the scaling variable is the total area $A$. The proper length of the throat measured in units of $r_+$ gives 
the number of e-foldings that $\Delta A$ decreases by from $\Delta A_\textrm{max}$ to $\Delta A(r_+)$ and this number 
diverges as the black hole approaches extremality. A corresponding scaling symmetry is found at low temperature 
in the SYK model \cite{Maldacena:2016hyu}.

\bn

As in all  theories with AdS symmetry the concept of a local energy scale  is a familiar feature \cite{Susskind:1998dq}. At the AdS boundary the energy scale diverges. It decreases as one moves deeper into the bulk. In AdS at a fixed value of 
$l_{ads}$ the local energy scale at radial coordinate $r$ can be defined as the temperature of a black hole whose horizon is at the radial coordinate $r$.

 Applying the same logic we may define the local energy scale along the throat to be the temperature $\tt$ of a ``virtual" black hole whose Rindler region begins at distance $\rho$ from the boundary. 
If $\rho $ denotes the start of the Rindler region of an actual black hole then $\tt(\rho)$ is the temperature of the black hole, i.e., $\tt(\rho) = T.$ Using the fact that the length of the throat for a black hole of temperature $T$ is 
$$\Delta \rho = r_+ \log{\beta/r_+} $$ we may write,
\be 
\tb(\rho) = r_+ e^{\rho/r_+}
\label{tb=exp}
\ee
where $\tb = 1/\tt.$  At the outer boundary of the throat,
\be 
\rho = 0 \ \ \rightarrow \ \  \tb = r_+  \ \ ;
\label{initial}
\ee
in the Rindler region, 
\be 
\rho = r_+ \log{\beta/r_+} \ \ \rightarrow \ \ \tb = \beta \ \ .
\ee
Thus we see that $\tt \equiv 1/ \tb$ varies a great deal throughout the throat.

Our proposal for generalizing \ref{old-proposal} is to simply replace \ref{old-proposal} by
\be 
\textrm{size} \sim  P\tb(r)
\label{s=pbtild},
\ee
in which size depends on both momentum and  position. This may simply be expressed as: 

\bn \it
Size equals momentum measured in units of the local energy scale.
\rm \\

Using \ref{tb=exp} the size-momentum relation takes the explicit form,
\be 
\textrm{size} \sim r_+e^{\rho/r_+} \ P.
\ee

\bn
The reason that  \ref{old-proposal} was sufficient for \S \ black holes is  simply 
because $\tb$ does not vary much between the boundary and the horizon.

\subsection{The Local Energy Scale and the Surface Gravity}

In Eq.~\ref{s=pbtild}, we conjectured that size and momentum are related by the position-dependent factor $\tilde{\beta}(r)$. We defined the (inverse) local energy scale $\tilde{\beta}(r)$ as the temperature that a black hole of the same charge  \emph{would} have if its mass were such that the event horizon lay at that value of $r$. In this subsection we will present an alternative perspective on the quantity $\tilde{\beta}(r)$ by providing a more directly physical definition. 

First consider the gravitational field, defined as the proper acceleration required to remain static
\begin{equation}
g = \alpha\biggl|_{\dot{r} = \dot{\Omega} = 0} = \partial_r \sqrt{f}.
\end{equation}
As we approach the black hole the gravitational field gets ever stronger, but it does not do so steadily. Instead, as we discussed in Sec.~\ref{sec:throatgeometry}, $g$ is gently increasing in the Newtonian region (due to the inverse square law), then constant in the throat ($g \sim 1/r_+$), then rapidly increasing in the Rindler region ($g \sim  1/(\beta \sqrt{f}$)), becoming infinite at the event horizon. 

On the other hand, the redshift factor $\sqrt{f(r)}$ steadily decreases as we approach the black hole. Far from the black hole it is of order one, in the throat it falls exponentially, and then in the Rindler region it falls even faster, approaching zero at the event horizon. 

The local energy scale is the product of these two factors, 
\begin{equation}
\tilde{\beta}(r)^{-1} = \sqrt{f(r)} g(r) =  \sqrt{f(r)} \partial_r \sqrt{f(r)} = \frac{1}{2} \partial_r f(r).
\end{equation}
In the throat, $g$ is constant, whereas $\sqrt{f}$ is decreasing, so the product is decreasing, $\tilde{\beta}^{-1} \sim \sqrt{f}/r_+$. In the Rindler region, the growth of $g$ cancels the fall in $\sqrt{f}$ and so the local energy scale is constant $\tilde{\beta}^{-1} \sim  \sqrt{f}/(\beta \sqrt{f}) \sim 1/\beta$.

A box may be held fixed near a black hole by suspending it from a rope, though this only works if someone is holding on to the other end. Physically, the local energy scale $\tilde{\beta}^{-1}$ is the force that must be exerted on the boundary-end of the rope to stop a unit-mass box falling into the black hole (ignoring the weight of the rope itself). Famously, this force is approximately constant for boxes in the Rindler region, and approaches the surface gravity as the box approaches the event horizon, $\kappa \sim \lim_{r \rightarrow r_+} \tilde{\beta}^{-1}(r)$. Since it is the surface gravity that controls the Hawking temperature, this explains the connection to the definition of Sec.~\ref{sec:localenergyscale}. 
\section{The Scrambling Time}

The following facts can be verified by explicit calculation.

\begin{enumerate}
\item As an infalling light-like trajectory, beginning at the boundary at $t=0$, passes through the throat the time and the value of $\tb$ are related by,
\be 
\tb = t + r_+.
\label{1}
\ee
\item A massless particle inserted at the boundary (top of the barrier at $\rho=0$) at $t=0$ will quickly attain a momentum $1/r_+.$ From \ref{initial} the initial value of  $\tb$ satisfies $\tb= r_+$ and the initial size is 
\be
s(0)=1.
\label{s(0)=1}
\ee

In the appendix we show that as it passes through the throat the momentum linearly increases,
\be 
P=1/r_+ + t/r_+^2.
\label{2}
\ee
Using \ref{1} we may write this in the form,
\be 
P = \tb/r_+^2 . 
\label{3}
\ee
\item
Combining \ref{3} and \ref{1} and using $\textrm{size} \sim \tb P$
gives a quadratic growth for size as the particle traverses the throat.
\be 
s(t)  \sim  P\tb  = \frac{(t+r_+)^2}{r_+^2}.
\label{4}
\ee
\item The particle arrives at the Rindler region after time $\beta.$ At that point the size is,
\be 
s(\beta) \sim \frac{\beta^2}{r_+^2}.
\label{5}
\ee

\end{enumerate}

Equation \ref{5} may be regarded as the initial condition when the particle enters the Rindler region. In the Rindler region the growth of the momentum is exponential \cite{Susskind:2018tei} and the energy scale hardly varies. Thus as the particle passes through the Rindler region the size grows according to,
\be 
s(\tau) \sim \frac{\beta^2}{r_+^2} e^{\tau}.
\label{6}
\ee\\

To find the scrambling time we set the size equal to the black hole entropy,
\be 
\frac{\beta^2}{r_+^2} e^{\tau_*} \sim S
\label{7}
\ee
giving,
\be 
\tau_* = \log{\Big( S\frac{ r_+^2}{\beta^2}\Big)},
\label{8}
\ee
or using \ref{DSoverS},
\be 
\tau_* = \log{\Big( (S -S_0 )\frac{ r_+}{\beta}\Big)}.
\label{9}
\ee

The final step is to use the first law of thermodynamics to relate $r_+/\beta$ to the increase of entropy due to the extra energy when the particle is absorbed into the black hole. Thus 
\be 
\tau_* = \log{\frac{(S -S_0 )}{\delta S}} , 
\ee
which exactly matches \ref{tau-scrmbl-2}.

To summarize, the scrambling time for a near-extremal Reissner-Nordstr{\"o}m black hole is smaller than might have been expected: it is proportional to $\log{\left\lbrace(S-S_0)/\delta S \right\rbrace }$ rather than $\log{\left\lbrace S/\delta S \right\rbrace }$. One possible explanation would have been that the extremal entropy is somehow frozen out of the scrambling process. But our analysis suggests a different reason. We saw that a particle falling through the throat is rapidly accelerated, so that its initial size is boosted  by a factor    $\beta^2/r_+^2$ by the time it enters the Rindler region.  This reduces the time required  for the size to grow to $S$. 
We therefore claim that the correct explanation for the reduced scrambling time of a charged black hole is not that the extremal degrees of freedom decouple, but instead that the size of an operator grows rapidly at early times, so by the time it starts its exponential growth the size is already large. The success in explaining the reduced scrambling time of charged black holes provides a non-trivial confirmation of the connection between size and momentum.\\

\section{Comment about GR=QM and SYK}\label{Sec: SYK}

 The real justification for the rules we have postulated must be micro-physical. Thus we turn to the SYK model \cite{Maldacena:2016hyu}, a model which has many feature in common with near-extremal black holes, but which has a precise micro-physical description.
The similarities between the two are well known and include the following:

\bi 
\item The overall energy scale $J$ of SYK  corresponds to the RN parameter $1/r_+.$ 
\item The dynamical boundary of SYK (described by the Schwarzian theory) corresponds to the boundary of the RN black hole, i.e.,  the top of the barrier where the throat meets the Newtonian region.
\item Acting with a single boundary fermion operator in SYK adds an energy $J$. This fits nicely with the fact that adding a particle at the top of the RN potential barrier adds energy $1/r_+.$
\item A single boundary fermion operator in SYK has size $1$, corresponding to our assumption that the initial size of the particle at the top of the barrier is also $1$.
\item
Until now it has not been possible to directly compare the results of size-momentum duality with calculations in the SYK theory, the reason being that the only calculations of size-growth were at infinite temperature  \cite{Roberts:2018mnp}.
However one of us (Streicher) and Xiaoliang Qi have recently carried out a finite temperature calculation
  using the OTOC method \cite{alex}.

  Here we will just quote the result,
\be 
s(t) = 1 + 2\left\{      \frac{J\beta}{\pi} 
\sinh
\left(\frac{\pi t}{\beta}  \right)      
\right\}^2.
\ee
We see that the initial size satisfies $s(0) =1,$ and that it grows quadratically as $s=J^2t^2$, in  agreement with \ref{4}. By $t=\beta$ the size has grown to approximately $J^2\beta^2$ and from thereon it grows exponentially as in \ref{6}. There appears to be close agreement between the SYK model and our analysis  based on size-momentum duality.

\ei

Finally we note that Eq.~\ref{scrambling-charged} was derived for the SYK model  \cite{Maldacena:2016hyu} on purely quantum mechanical grounds, without any assumption of a dual geometry. It  is extremely interesting that a whole class of very generic quantum systems reproduce a formula whose interpretation is both geometric and gravitational: the existence of a long throat (geometry)  through which a particle will accelerate as it falls toward the horizon (gravity). This is another example of ``GR=QM'', i.e.~the view that the origins of gravity are to be found in the generic behavior of complex quantum systems \cite{Susskind:2017ney}.

  \section*{Acknowledgements}
  
  We are grateful to Juan Maldacena, Steven Shenker, and Douglas Stanford for discussions. This research was supported by the John Templeton Foundation and by FQXi grant RFP-1606 (AB), by NSF grant PHY-1720397 (HG), by the Simons Foundation (AS), by NSF Award Number 1316699 (LS), by the Icelandic Research Fund under grant 163422-053 and the University of Iceland Research Fund (LT), and by the Stanford Institute for Theoretical Physics (YZ).

\sc
\appendix
\section{Particle Equation of Motion }

The radial part of the metric \ref{RN-metric} may be written in the form,
\be 
ds^2 = -f(r) dt^2 + d\rho^2.
\ee
The Lagrangian for a point mass moving in this metric is,
\be 
\CL = -m\sqrt{f(r) - \dot{\rho}^2}.
\ee
The canonical momentum conjugate to $\rho$ is,
\bea 
P\eq \frac{\partial \CL}{\partial \dot{\rho}} \cr
\eq \frac{m\dot{\rho}}{\sqrt{f(r) - \dot{\rho}^2}}.
\eea
The  conserved Hamiltonian is
\be 
H \equiv P \dot{\rho} - \CL = \frac{mf}{\sqrt{f(r) - \dot{\rho}^2}}.
\label{ham}
\ee
The force on the particle is,
\bea
F \eq \frac{\partial \CL}{\partial {\rho}} \cr \cr
\eq -\frac{m \ \partial_{\rho}f}{2\sqrt{f-\dot{\rho}^2}}\cr \cr
\eq   -\frac{m \ \partial_{r}f}{2\sqrt{f-\dot{\rho}^2}} \ \frac{dr}{d\rho} \cr \cr
\eq  -\frac{m \ \partial_{r}f}{2\sqrt{f-\dot{\rho}^2}} \ \sqrt{f}
\eea

Using \ref{ham} we get
\be 
F=  -\frac{\partial_r f}{2\sqrt{f}}   H
\ee

So far this is general. Now plugging in the metric \ref{RN-metric}  (we use the extremal form with $r_+ = r_-$)  the force is given by,
\be 
F=-\frac{r_+ H}{r^2}.
\ee

In the long throat $r$ is almost constant and equal to $r_+.$ Furthermore, a particle dropped from the top of the potential barrier has (conserved) energy $H = 1/r_+.$ It follows that the force on the particle as it traverses the throat is constant and equal to,
\be 
F \sim -\frac{1}{r_+^2}.
\ee
Finally we use the equation of motion $dP/dt = F$ to find,
\be 
|P|=\frac{t}{r_+^2} + const
\ee
in agreement with \ref{2}.

\end{document}